\documentclass[prl,twocolumn,amsmath,amssymb,amsfonts,superscriptaddress,longbibliography]{revtex4-1}

\usepackage{graphicx}
\usepackage{hyperref}
\hypersetup{colorlinks=true}

\usepackage{amsmath}
\DeclareMathAlphabet{\mathpzc}{OT1}{pzc}{m}{it}
\newcommand{\dnu}{\mathpzc{W}}

\newcommand{\nair}{\textrm{air}}
\newcommand{\npl}{\textrm{pl}}
\newcommand{\nouter}{\textrm{outer}}
\newcommand{\nDC}{\textrm{DC}}
\newcommand{\ndecay}{\textrm{decay}}

\newcommand{\be}{\begin{equation}}
\newcommand{\ee}{\end{equation}}

\begin{document}

\title{Detection of Surface Waves During Femtosecond Filamentation}


\author{Travis Garrett}
\affiliation{Air Force Research Laboratory, Directed Energy Dicrectorate, Albuquerque, NM 87123, USA}

\author{Anna Janicek}
\affiliation{Air Force Research Laboratory, Directed Energy Dicrectorate, Albuquerque, NM 87123, USA}

\author{J.~Todd Fayard II}
\affiliation{Department of Physics and Astronomy, Mississippi State University, Mississippi State, MS 39762, USA}

\author{Jennifer Elle}
\affiliation{Air Force Research Laboratory, Directed Energy Dicrectorate, Albuquerque, NM 87123, USA}


\begin{abstract}

Ultrashort pulsed lasers (USPL) can produce thin columns of plasma in air via  
femtosecond filamentation, and these plasmas have been found to generate 
broadband TeraHertz (THz) and Radio Frequency (RF) radiation.
A recent theory argues that the currents driven at the boundary of the plasma 
excite a Surface Plasmon Polariton (SPP) surface wave 
(in particular a Sommerfeld-Goubau wave given the cylindrical symmetry), 
which proceeds to detach from the end of the plasma 
to become the RF pulse. 
We have performed near-field measurements of these plasmas with a 
D-dot probe, and find an excellent agreement with this theory.
The radial field dependence is precisely fit by a Hankel 
function, with an outer length scale in agreement with plasma 
conductivity and radius, and a measured longitudinal drift in 
frequency maxima closely matches both SPP simulations 
and analytic expectations.  

\end{abstract}

\maketitle

\section{Introduction}
Surface waves can be excited across a spectrum of environments, 
including solitonic shallow water gravity waves 
\cite{kordeweg1895change,zabusky1965interaction,ablowitz1979evolution}, 
Rayleigh and Love surface acoustic waves 
(from microfluidic to planetary scale)
\cite{rayleigh1885waves,love1911some,delsing20192019}, 
Dyakonov, Bloch, and Tamm electromagnetic surface waves 
along special dielectrics
\cite{d1988new,yeh1977electromagnetic,polo2011surface,takayama2017photonic,jahani2016all}, 
and surface plasmon polaritons \cite{ritchie1957plasma,pitarke2006theory} 
on the boundaries of conducting substrates 
with free electron number density $n_e$.
Surface plasmon polaritons are composed of an electromagnetic  
wave coupled to a charge density wave on the boundary of 
the metal, semiconductor or plasma region, 
and can exist under a cutoff frequency $\omega_s$, 
which for a flat plane boundary equals $\omega_s = \omega_{\textrm{pl}}/\sqrt{2}$,
where $\omega_{\textrm{pl}}$ is the plasma frequency.

SPPs are typically considered along the boundaries 
of metals and semiconductors in the optical 
and infrared bands, where they can be driven 
by Otto or Kretschmann coupling 
\cite{otto1968excitation,kretschmann1968radiative},
but they can be excited along the boundary of a plasma as well. 
An Ultra Short Pulse Laser (USPL) can generate volumes of plasma  
in a unique way: given sufficient power Kerr self-focusing 
will boost the intensity of a standard 800nm, 50fs, Ti:Sapphire pulse 
propagating in air to roughly $5 \times 10^{17}$ W/m${}^2$, 
at which point strong field ionization 
\cite{popruzhenko2008strong}
creates a plasma of density roughly $n_e \sim 10^{22}$ m${}^{-3}$
and radius $r_{\textrm{pl}} \sim 100$ $\mu$m. 
USPL filamentation thus generates thin columns 
of plasma that grow longer at the speed of light
\cite{braun1995self,theberge2006plasma,couairon2007femtosecond}, 
which have a wide array of potential applications 
\cite{kasparian2008physics,chin2012advances,kudyshev2013virtual,
rostami2016dramatic,kasparian2012ultrafast}. 

In addition to creating the plasma, the 
femtosecond pulse also endows the plasma with initial 
current densities. 
A $\textbf{J} \times \textbf{B}$ Lorentz force 
drives a direct longitudinal current density  
in the second half of the laser pulse
($J_z$ for a pulse propagating in $\hat{z}$), 
which is closely linked to 
broadband single color filament THz 
\cite{cheng2001generation,sprangle2004ultrashort,d2007conical,amico2008forward}
(with an associated low frequency RF tail).
Zhou et al \cite{zhou2011measurement} noted that transverse 
current densities ($J_x$ for a linear $\hat{x}$ polarized laser pulse, 
that evolves into $J_r$ with collisions) 
are also generated by the laser pulse, 
and they argue that the resulting plasma wake trailing 
behind the laser pulse will also excite RF radiation.

Stimulated by far field measurements of broadband RF 
from filamentation
\cite{forestier2010radiofrequency,englesbe2018gas,janicek2020length,
mitrofanov2021coherently,englesbe2021ultrabroadband},
several years ago our group developed a quantitative 
model of RF generation due to the transverse plasma currents 
\cite{garrett2021generation}.
Particle In Cell (PIC) simulations 
\cite{birdsall2004plasma,peterkin2002virtual}
confirmed that a 
plasma wake is established on the outer boundary of the plasma; 
the magnitude of the associated $J_r$ current is limited 
by electron neutral collisions at higher pressures, 
and by the build up of an electrostatic well at lower pressures 
(e.g. 1 Torr).

Larger scale continuum Drude simulations 
\cite{okoniewski1997simple,teixeira2008time}
then lead to a surprise: 
the presence of the plasma wake current pulse $J_r$ on the 
outer boundary of the plasma acts as an antenna 
and invariably excites a SPP (see also \cite{chen2012optical}).
Furthermore the RF frequencies of this SPP are far 
below the cutoff at $\omega_s \sim 1$ THz, and they thus travel 
with a group velocity close to $c$.
The plasma wake currents, translating at $c$ behind the laser 
pulse, thus co-propagate coherently with and steadily  
amplify the SPP surface wave at the leading edge of the plasma column. 
Simulations also show that the surface wave detaches efficiently from 
the end of the plasma column (end fire radiation 
\cite{andersen1967radiation,stegeman1983excitation}) to become the forward 
directed pulse of RF. 

The magnitude and spatial profile of the RF produced by this 
plasma wake surface wave model was 
found to accurately match far field broadband 
horn laboratory measurements \cite{garrett2021generation}.
In this work we revisit the theory with near-field 
D-dot measurements in the vicinity of the filamentation plasma.
We find strong evidence of the Sommerfeld-Goubau SPPs 
predicted by the theory, with an excellent fit 
to a Hankel function radial profile and the 
$r_{\textrm{outer}}$ length scale. 
We have also detected a longitudinal frequency 
drift in the SPPs that closely matches 
new axisymmetric Drude simulations, 
and new theoretical derivations for the phase velocity 
and resistive attenuation length scales. 
The measurement and characterization of these surface waves thus provides 
a new window into the physics of filamentation plasmas 
and for chirped pulse laser science in general.

\section{Theory}

The theory developed in \cite{garrett2021generation} was developed 
through exploratory simulations at a variety of length scales, 
and then simplified into analytic approximations.
Small scale PIC simulations that resolve the 800 nm wavelength  
and make use of the ionization rate 
$\dnu$ given in \cite{popruzhenko2008strong} 
reproduce the electron velocity distributions given in 
\cite{zhou2011measurement} for linearly and circularly 
polarized laser pulses 
(and the large net transverse current for two color 
systems \cite{kim2007terahertz}).
In the simplest PIC simulations the electrons are only accelerated 
by the remainder of the laser pulse, 
which yields the highest residual velocities for electrons born with 
instantaneous $|E| = 0$, but one can also include interactions 
with the parent ion which leads to additional heating 
(see also \cite{corkum1993plasma}).

We then perform larger PIC simulations of transverse slices 
of this plasma column where the initial electron velocities 
are set by the previous simulation. 
With 50 mJ of energy, the 50 fs pulses are in the multifilamentary 
regime \cite{berge2004multiple}, and we approximate the bundle of 
filament plasmas as one larger plasma column with radius $r_{\textrm{pl}}=500$ $\mu$m.
At atmospheric pressure 
the tail of the distribution is at about $K_{\textrm{tail}}=4$ eV, and with an 
electron-neutral collision frequency $\nu$ of $5$ THz these diffuse 
out to roughly $r_{\Delta} = 30$ $\mu$m over 100 ps, with an 
effective late time diffusion velocity of 
$v_{\textrm{eff}} = 7\times10^4$ m/s. An analytic approximation for 
the radial current density $J_r$ is given by: 
$J_r = \epsilon_0 v_{\textrm{eff}} K_{\textrm{tail}} / r_{\Delta}^2$,
which is about $3000$ A/m${}^2$ at 10 GHz. 

\begin{figure}[h!]
  \centering
  \includegraphics[width=0.5\textwidth]{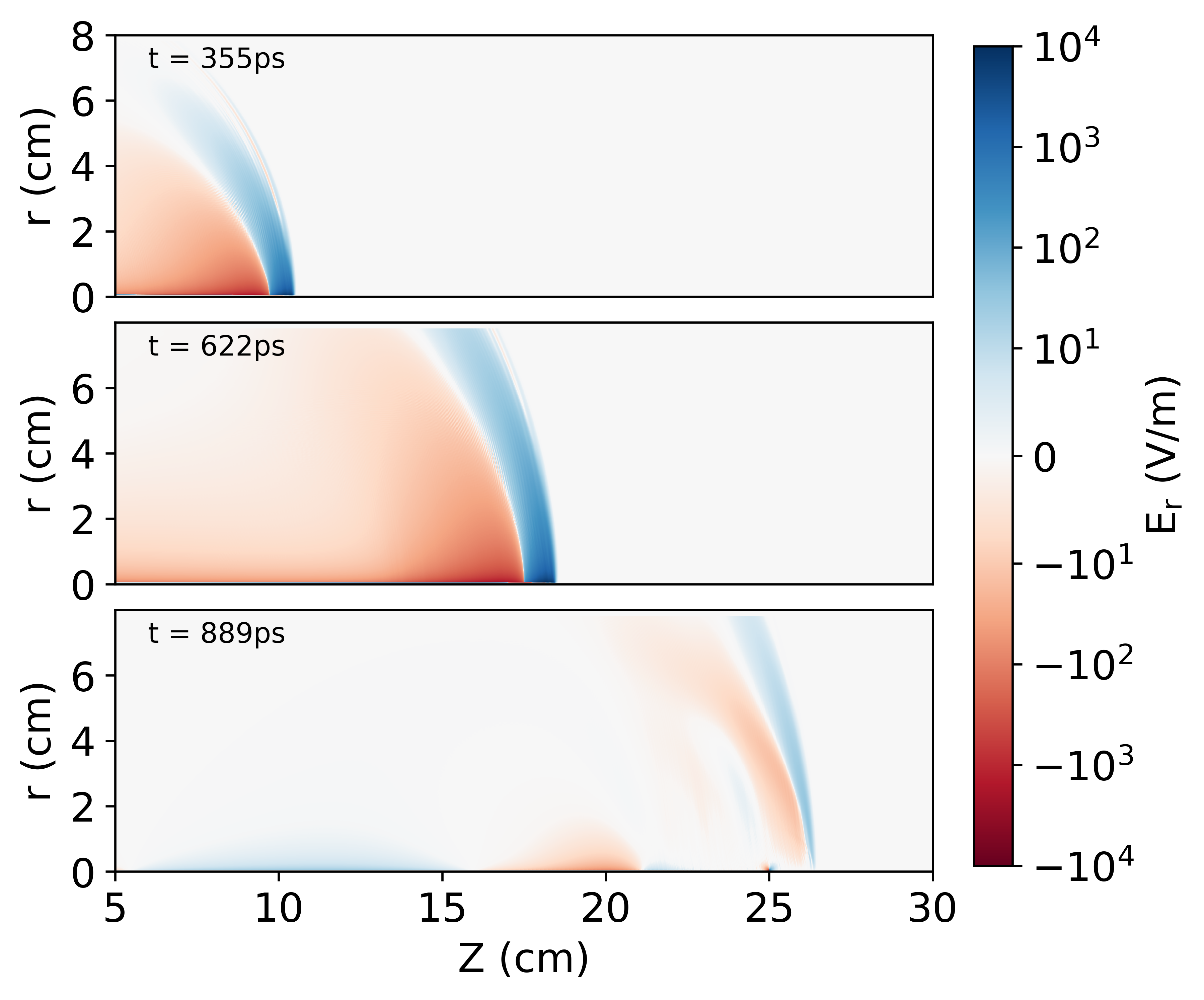}
  \caption{ Log plots of $E_r$ for an axisymmetric  
  Drude electromagnetic simulation at 3 points in time, 
  with the SW near $z = 10$ cm and $z = 18$ cm at the 
  top and middle, and free RF radiation shortly after detachment from the 
  $25$ cm plasma column on the bottom. The plasma column 
  has a central number density of $n_e = 10^{23}$ m${}^{-3}$, and has been given 
  a $\sin^2$ envelope in the $\hat{z}$ direction 
  (inspired by \cite{janicek2023measuring}), which causes 
  the SW to both decay and smoothly detach 
  over the final $5$ cm with minimal upstream 
  reflection (in constrast to the substantial reflection 
  seen in previous uniform $\hat{z}$ density simulations 
  \cite{garrett2021generation}).}
  \label{sim_pics}
\end{figure}

The presence of this broadband radial current pulse $J_r$ on the outer boundary 
of the plasma column was found to excite a broadband SPP in  
axisymmetric Drude simulations \cite{garrett2021generation}; 
Fig. \ref{sim_pics} shows an updated simulation. 
This surface wave is well approximated by the transverse magnetic Sommerfeld-Goubau solution 
\cite{sommerfeld1899ueber,goubau1950surface,stratton2007electromagnetic,
orfanidis2002electromagnetic,pfeiffer1974surface,wang2006dispersion}.  
At a particular angular frequency $\omega$ 
the radial component of the electric 
field $E_r$ outside of the plasma is given by:
\begin{equation}
	E_r (r,z,t) = -\frac{\pi r_{\npl} E_0}{2 r_{\nouter}}e^{i(\omega t - hz)}H_1^{(1)}\left(\gamma_{\nair} r\right),
	\label{eqn_Sommerfeld_Goubau}
\end{equation}
where $E_0$ is the amplitude of the wave at the surface of the plasma 
at $r_{\textrm{pl}}$ (about $10^4$ V/m at atmospheric pressure), the 
external $\gamma_{\nair}$ is given by  
$\gamma_{\nair}^2 = k_0^2 \varepsilon_{\nair} - h^2$,
with $\varepsilon_{\nair} = \epsilon_{\nair} / \epsilon_0$,  
free space wave number $k_0=\omega/c$, complex 
wavenumber $h$ which encodes both the SPP wavelength and attenuation length scale 
($L_{\ndecay} = 1/\textrm{Im}(h)$), 
and $H_1^{(1)}$ is a Hankel function of the first kind with order 1, 
with outer length scale $r_{\nouter}=1/|\gamma_{\nair}|$.
For radii $\hat{r}$ between $r_{\npl}$ and $r_{\nouter}$
the Hankel function is approximately $H^{(1)}_1 \simeq -2 r_{\nouter}/(\pi r)$, 
and for radii larger than $r_{\nouter}$ it transitions to falling 
off exponentially: we fit the experimental D-dot data 
at a variety of frequency bands and radii to this predicted profile.

\begin{figure}[h!]
  \centering
  \includegraphics[width=0.5\textwidth]{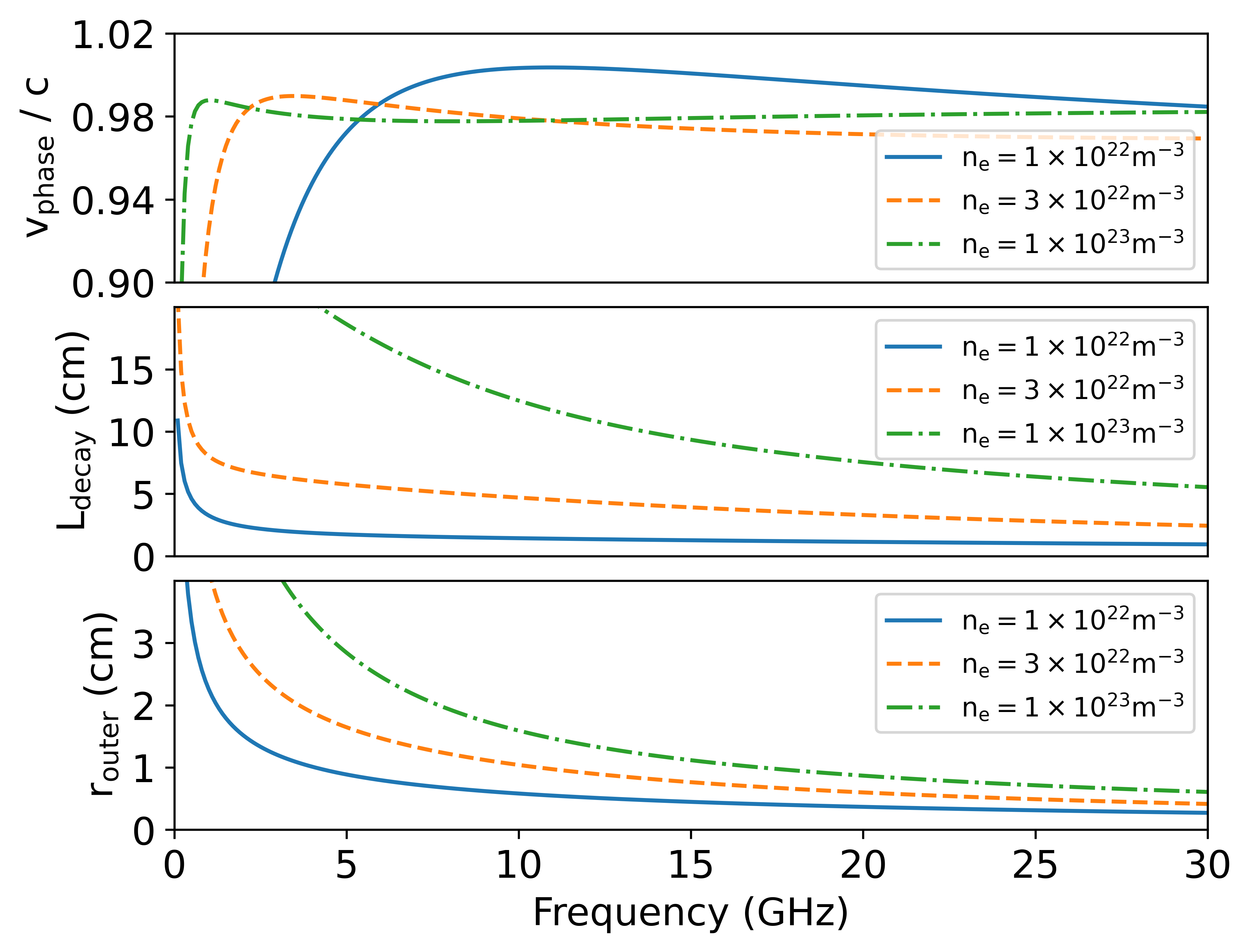}
  \caption{ 
  Plots of the phase velocity, decay length, and outer length scale 
  as a function of SPP frequency as calculated by (\ref{iterative_solve}), 
  for a plasma column with radius $r_{\npl}=0.5$ mm, 
  collision frequency $\nu=5$ THz, and density $n_e$ 
  ranging from $10^{22}$ to $10^{23}$ m${}^{-3}$.
  The phase velocity slowly climbs towards $c$ as the frequency 
  decreases (as for planar SPPs), and then plunges when the skin depth becomes
  comparable to $r_{\npl}$. $L_{\ndecay}$ and $r_{\nouter}$ both 
  increase for lower frequencies, with $L_{\ndecay}$ falling 
  rapidly with decreasing $n_e$, and $r_{\nouter}$ more slowly. }
  \label{theory_pics}
\end{figure}

The $r_{\nouter}$ length scale depends on $h$, which is found by matching 
the internal $E_z$ and $H_{\phi}$ fields (having $J_0$ and $J_1$ Bessel function dependences respectively) to 
the external fields (described by $H^{(1)}_0$ and $H^{(1)}_1$ Hankel functions) at the plasma column outer 
radius $r_{\npl}$:
\be
\frac{\gamma_{\nair} H^{(1)}_0(\gamma_{\nair}r_{\npl})}{\varepsilon_{\nair} H^{(1)}_1(\gamma_{\nair}r_{\npl})} = 
\frac{\gamma_{\npl} J_0(\gamma_{\npl}r_{\npl})}{\varepsilon_{\npl} J_1(\gamma_{\npl}r_{\npl})},
\label{matching_condition}
\ee
with $\gamma_{\npl}^2 = k^2_0 \varepsilon_{\npl} - h^2$ and relative plasma 
permittivity $\varepsilon_{\npl}$ \cite{orfanidis2002electromagnetic} 
(and setting $\varepsilon_{\nair}=1$ from here on).

Sommerfeld originally solved (\ref{matching_condition}) by using asymptotic approximations  
for the Bessel and Hankel functions and then developing 
an iterative method \cite{sommerfeld1899ueber} to find $h$,
and recent work has shown that those approximations can also yield 
a direct solution via the Lambert $W$ function
\cite{jaisson2014simple,mendoncca2019electromagnetic}.  
We used this direct Lambert $W$ method to estimate 
$h$ in \cite{garrett2021generation}, 
but to get more accurate values for this work 
we have dropped the approximations and 
switched to an iterative method that uses the 
full Bessel and Hankel functions (see \cite{orfanidis2002electromagnetic}).
In particular, we include both the low and high frequency terms in the plasma 
permittivity:
\be
\varepsilon_{\npl}(\omega) = 1 + \frac{\omega_{\npl}^2}{i \omega \nu - \omega^2},
\label{eps_plasma}
\ee
since $\omega \epsilon_0$ is not too much smaller than $\sigma_{\nDC} = \epsilon_0 \omega^2_{\npl}/\nu$.
At lower frequencies 
the skin depth $\sqrt{2/\sigma_{\nDC}\mu\omega}$ can be comparable 
to the plasma radius $r_{\npl}$, 
and so we also do not use the $J_0/J_1 = i$ approximation.
We thus rearrange the full (\ref{matching_condition}) equation 
to give the $N+1$ iteration for $\gamma_{\nair}$: 
\be
\gamma_{\nair}^{N+1} =  
\frac{\gamma_{\npl}^N H^{(1)}_1(\gamma_{\nair}^N r_{\npl}) J_0(\gamma_{\npl}^N r_{\npl})}
{\varepsilon_{\npl} H^{(1)}_0(\gamma_{\nair}^N r_{\npl}) J_1(\gamma_{\npl}^N r_{\npl})},
\label{iterative_solve}
\ee
with $\gamma_{\npl}^N = \sqrt{(\gamma_{\nair}^N)^2+(\varepsilon_{\nair}(\omega)-1)k^2_0}$, 
an initial $N=1$ guess of $\gamma_{\nair}^{N=1} = 0.5 k_0$, 
and $h = \sqrt{k_0^2 - \gamma_{\nair}^2}$ once the iteration has converged 
(typically 10-30 steps - see \cite{mendoncca2019electromagnetic}
for further discussion on convergence criteria).
For a plasma with an estimated values $r_{\npl}=500$ $\mu$m, $\nu \sim 5$ THz,
$n_e \sim 3 \times 10^{22}$ m${}^{-3}$ (giving $\sigma_{\nDC} \sim 170$ S/m), 
and a surface wave frequency of 10 GHz, (\ref{iterative_solve}) yields 
$h \simeq 214 + 21i$, which gives $r_{\nouter} \sim 1.0$ cm, 
and a longitudinal $1/e$ attenuation length scale of $L_{\ndecay} \sim 4.7$ cm. 
We note that with longer laser pulses the electron density $n_e$ can grow 
much higher, especially as collisional heating and ionization become important 
\cite{thornton2024boosting}, 
which leads to higher conductivity, a larger plasma radius and better surface wave 
propagation and growth.

We base our theory on the Sommerfeld-Goubau solution (\ref{eqn_Sommerfeld_Goubau}), 
but additional dynamics occur in the surface wave 
simulations (see Fig. \ref{sim_pics}) that are not easily captured by analytic approximations.
The analytic solutions assume a Heaviside step function profile in the conductor 
electron density, which is a good approximation for optical SPPs on 
metal substrates, but less so for the boundaries of plasmas. 
However, in simulation when we switch from a step function to a radial electron 
density gradient at the outer boundary of the plasma we get qualitatively 
similar SPPs, with the amplitude $E_0$ of these gradient-boundary waves somewhat suppressed, 
and the spatial wave length moderately increased.

We also observe that the plasma wake radial current directly excites some 
free radiation in addition to the surface wave: this lower amplitude free radiation can 
be seen to slowly pull ahead of the surface wave over long simulations
(and can be seen in Fig. \ref{sim_pics} given the log scale). 
We suspect that the D-dot probe is also detecting this prompt free 
radiation in the experiments.

The original simulations performed in \cite{garrett2021generation} 
used plasma columns with constant density in the longitudinal $\hat{z}$ direction, 
but given the experimental measurements in 
\cite{janicek2023measuring} we have switched to a more realistic 
tapered $n_e(z) = n_{e,max}\sin^2(\pi z/L_{pl})$ profile, 
with low plasma densities at the beginning and 
end of the column. 
The use of this tapered profile results in modulation 
of the SPP amplitude that closely mirrors the near field experimental data. 
It also helps to explain the absence 
of higher frequency far field RF in experiments such as \cite{englesbe2018gas}
(given the presence of high amplitude, high frequency content in $J_r$), 
as higher frequency SPPs are more strongly attenuated in low density regions
(see Fig. \ref{theory_pics}). 
We also find a frequency dependent drift in the simulated SPP peak amplitudes that 
closely matches the longitudinal D-dot measurements. 
We note that the dispersion of SPPs in a cylindrical geometry 
is more complex than on a plane, 
with the group velocity of low frequency waves 
dropping rapidly instead of converging to $c$ 
\cite{pfeiffer1974surface,wang2006dispersion}, 
as can be seen in Fig. \ref{theory_pics}.
On a filamentation plasma where Kerr focusing 
and plasma defocusing can compete to form islands 
of high density plasma this can cause the lowest frequency SPPs 
to become trapped on the high density islands.

\section{Experimentation}

\begin{figure}[hbt!]
  \centering
  \includegraphics[width=.9\linewidth]{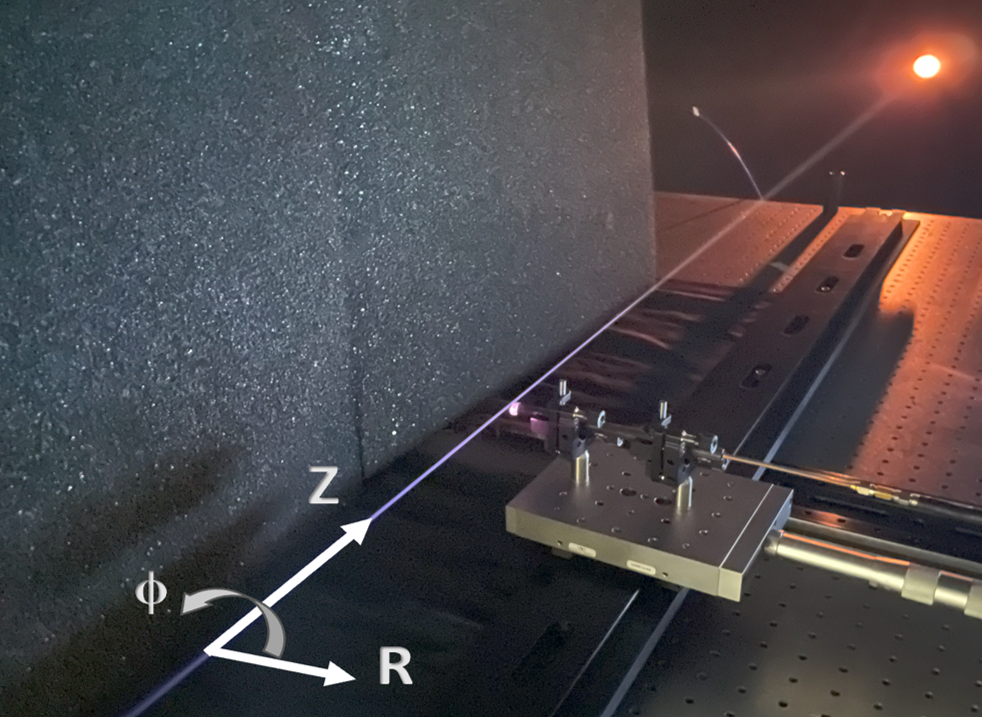}
  \caption{The 800nm laser plasma column was generated with the aid 
           of lens assisted filamentation to conserve laboratory space.  
           The D-dot probe is located on an $\hat{r}$ direction translation stage 
           and is swept in the $\hat{z}$ direction using an optical rail. }
  \label{setup1}
\end{figure}

\begin{figure}[hbt!]
  \centering
  \includegraphics[width=.9\linewidth]{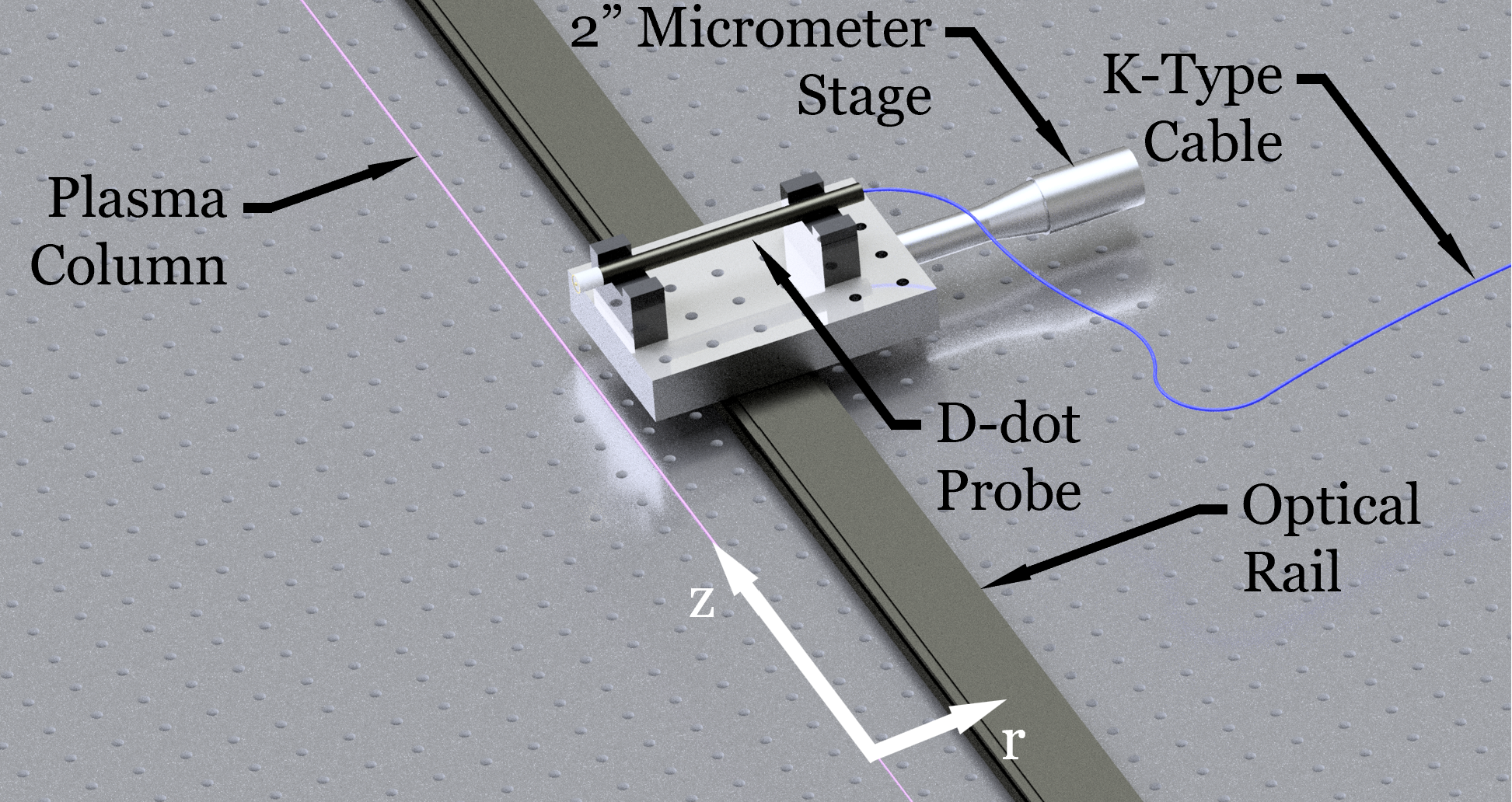}
  \caption{The above rendering shows the components utilized during experimentation.  
            The Prodyn FD-5C2 D-dot probe is mounted atop a 2 inch micrometer stage 
            which moves along the $\hat{r}$ axis.  
            Additionally, the stage is mounted to a 72.5 inch optical rail; 
            this allows for movement along the $\hat{z}$ axis.  
            The plasma column propagates along the positive $\hat{z}$ axis as shown above.  
            K-type coaxial cable was used to connect to the oscilloscope, 
            and microwave foam was placed parallel to the plasma to reduce signal reflections. }
  \label{setup2}
\end{figure}

To investigate the near-field region of a femtosecond plasma, 
an 800 nm wavelength 50 mJ laser was used with a 50 fs pulse duration and a 10 Hz repetition rate. 
Plasma formed near the geometric focal point of a 3 m focusing optic
(leading to somewhat higher $n_e$).  
Fig. \ref{setup1} shows the visible florescence of the plasma column 
which spans approximately 18 cm across the top of an optical bench.

A Prodyn FD-5C2 D-dot probe (bandwidth of 2-40 GHz) was used to measure 
the electric field generated by the plasma column. 
From preliminary experiments, the strongest signal was received by the D-dot probe 
when placing the probe tip normal to the direction of plasma propagation (see Fig. \ref{setup1}).  
The probe was mounted to a 2 inch micrometer stage (movable in the $\hat{r}$ direction) 
in the same plane as the plasma column.  
The stage was attached to two consecutive 33 inch optical rails mounted along the direction of propagation.  
This allowed for the probe to be positioned both along the $\hat{z}$ direction 
of the plasma and at various radius locations in the $\hat{r}$ direction.  
By mounting the probe in such a configuration, the plasma column was able to be examined 
in increments of 1/16 inch in the $\hat{z}$ direction and 0.001 mm in the $\hat{r}$ direction.  
The laser was carefully aligned through two irises to ensure the D-dot probe 
would remain at a fixed radius from the plasma column while moving along the z axis.  
An index card flush with the D-dot was periodically ablated by the plasma column at various $\hat{z}$ distances 
to confirm the distance between the probe tip and the plasma column remained constant.  
The experimental setup is depicted by the rendering in Fig. \ref{setup2}. 

Data was collected by selecting a stationary radius $r$ and longitudinally scanning 
in the $\hat{z}$ direction with the D-dot probe; r values of 0.5, 0.75, 1.0, 1.5, and 2.0 cm 
were probed at 28 locations along the $\hat{z}$ axis ranging from 248.9 cm to 320.0 cm 
from the focusing optic.  
Steps along the $\hat{z}$ axis were taken in 2.54 cm increments.  
Additionally, radial data was collected at the fixed longitude location 
of highest signal ($z$ = 294.6 cm) for 19 evenly spaced radii between 0.25 cm and 5.0 cm from the plasma surface
(note that we later shift our z coordinate system to begin with  
of plasma formation, so that $z$ = 295 cm becomes $z$ = 18 cm). 
For both experiments laser output energy was measured at the beginning and end of each test.  
The signal received by the D-dot probe was digitally captured using a Tektronix DP077002SX 33 GHz oscilloscope 
which limited the upper frequency limit of the measurements.  
The D-dot probe was connected to the oscilloscope using a K-type (40 GHz) cable.

\section{Results}

After collecting the raw data in the Tektronics digital oscilloscope, 
a Fast Fourier Transform (FFT) was applied to the time dependent signals  
to obtain the frequency content within the 2-33 GHz microwave regime. 
The noise floor from the laboratory environment was also recorded, 
and the resulting noise FFT subtracted from the signal FFTs.  
We note that developing an accurate, frequency dependent calibration 
of the D-dot probe (see e.g. \cite{englesbe2018gas}) 
is non-trivial, and we leave it for future work. 
As the D-dot amplitudes are uncalibrated, 
the measured values of the electric fields 
at different frequencies should only be considered in a relative context.

\begin{figure}[h!]
  \centering
  \includegraphics[width=0.5\textwidth]{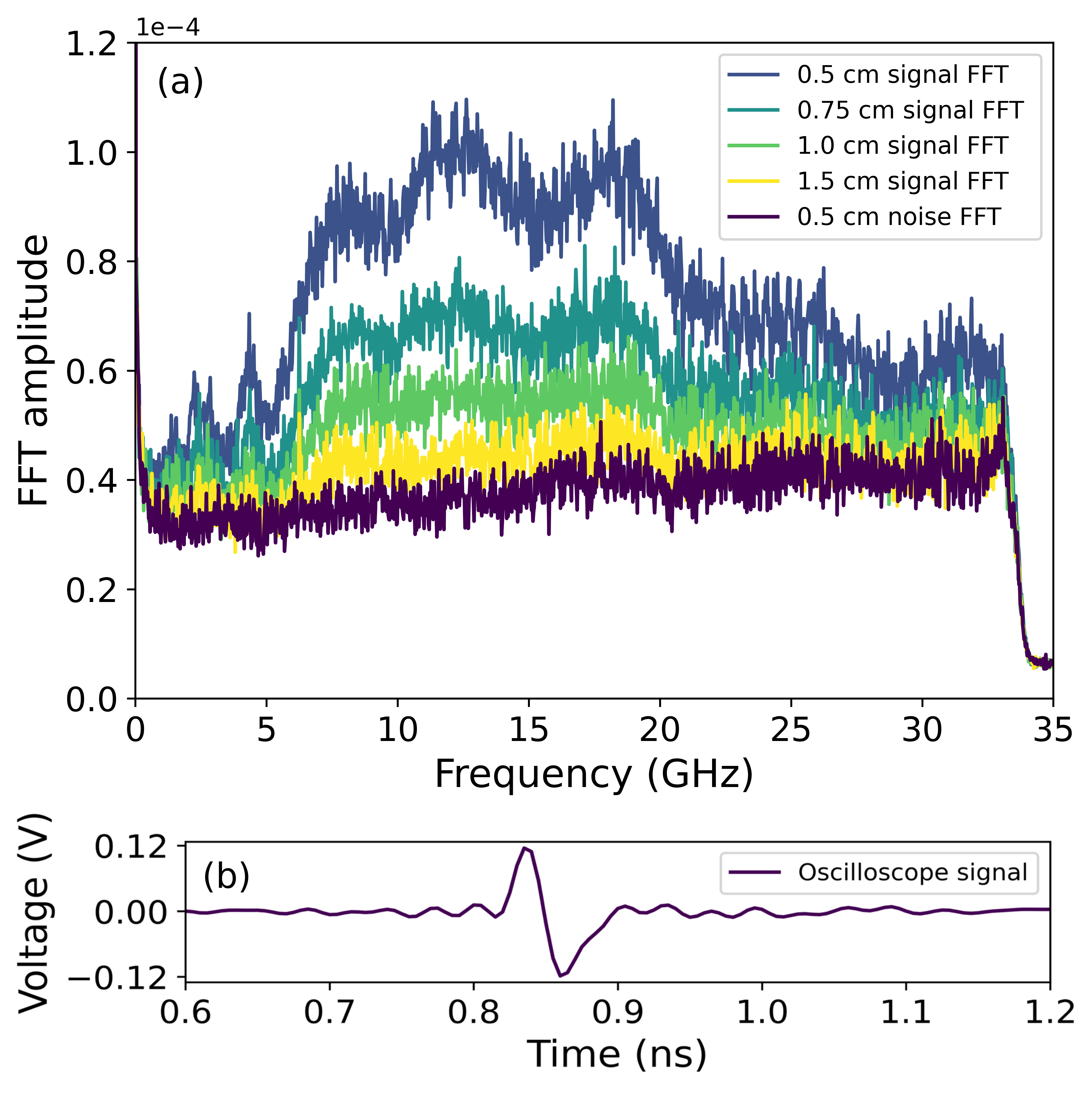}
  \caption{(a) FFTs of measured electron field strength measured at 
  $z = 18$ cm (near the peak) for radial separations ranging 
  from 0.5 cm to 1.5 cm, along with the measured noise spectrum 
  taken when no filamentation is occurring. The near field D-dot probe 
  has not been calibrated for this experiment, but the broadband 
  signal resembles the far field data taken in \cite{englesbe2018gas}.
  In all frequency bins the amplitude of the signal above the noise 
  floor falls off rapidly with increasing distance.
  (b) An example time domain signal recorded by the oscilloscope 
  at a separation of 0.5 cm. }
  \label{fft_plots}
\end{figure}

\begin{figure}[hbt!]
  \centering
  \includegraphics[width=0.5\textwidth]{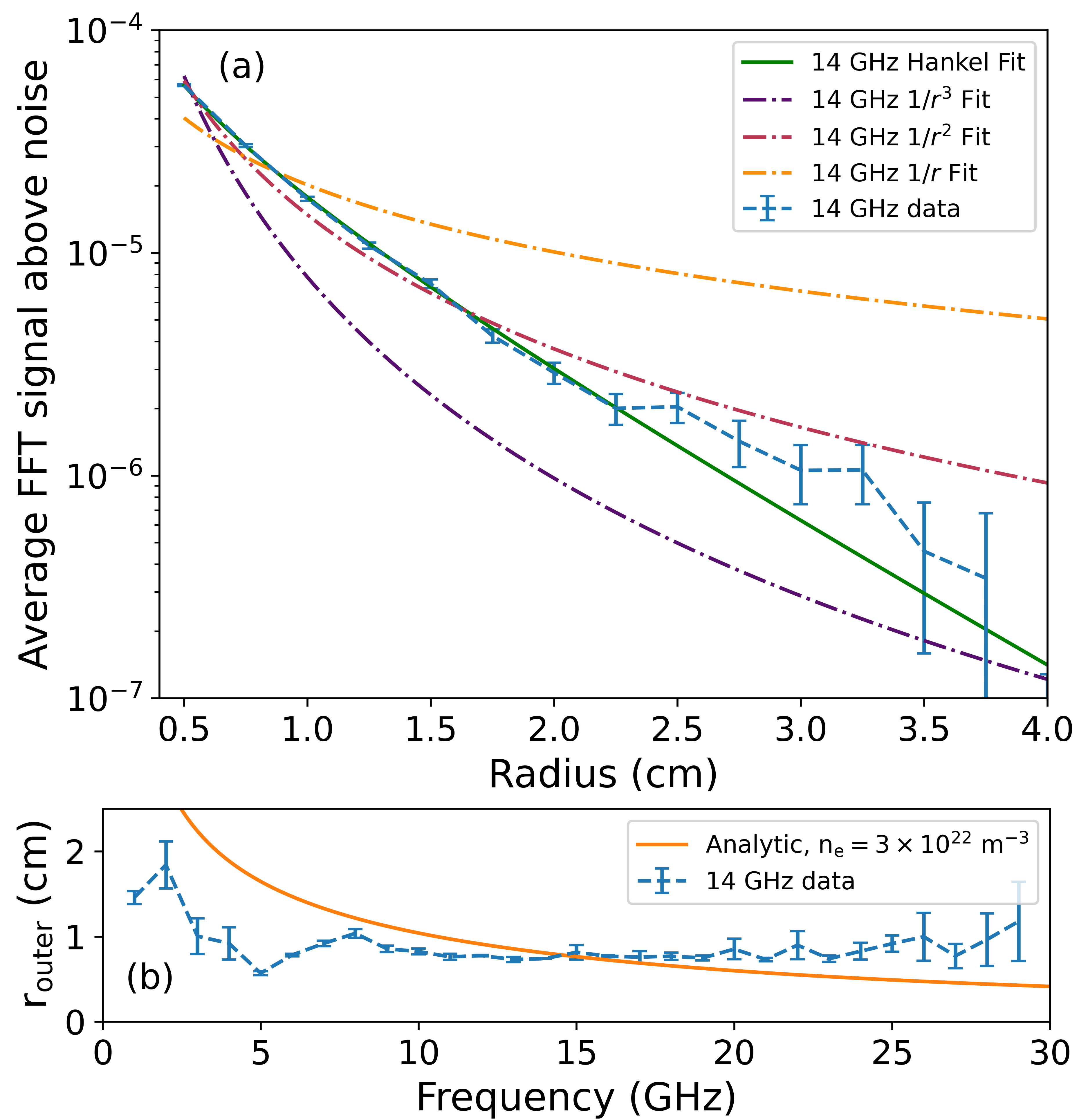}
  \caption{ (a) Log scale plot of the 14 GHz bin data (collected at $z$ = 18 cm) 
            for radii between $r$ = 0.25 cm and 3.75cm.
            The radial data in all frequency bins is well fit 
            by the expected Hankel function dependence (\ref{eqn_Sommerfeld_Goubau}), 
            with typical $R^2$ values of $0.99$, and the 14 GHz bin yields the 
            best fit, with $R^2 = 0.9993$.
            Visually the Hankel fit is very good out to a distance of about 2 cm,
            much better than the naive $1/r$, $1/r^2$, and $1/r^3$ profiles,  
            and then it slightly underpredicts the measured signal at larger radii. 
            We suspect that this is due to the presence of 
            free radiation, in addition to the bound SPP wave, 
            which the simulations also generically predict to be excited 
            (see Fig. \ref{sim_pics}).
            (b) Comparison of the $r_{\nouter}$ values from 
            Hankel function fits as a function of frequency, 
            compared to the analytic expectation from 
            (\ref{iterative_solve}) with an electron density 
            of $n_e = 3\times10^{22}$ m${}^{-3}$. The overall magnitude is 
            quite close, although the slope of the experimental 
            data is flatter than the theoretical expectation. }
  \label{hankel_fits}
\end{figure}

\begin{figure}[h!]
  \centering
  \includegraphics[width=0.5\textwidth]{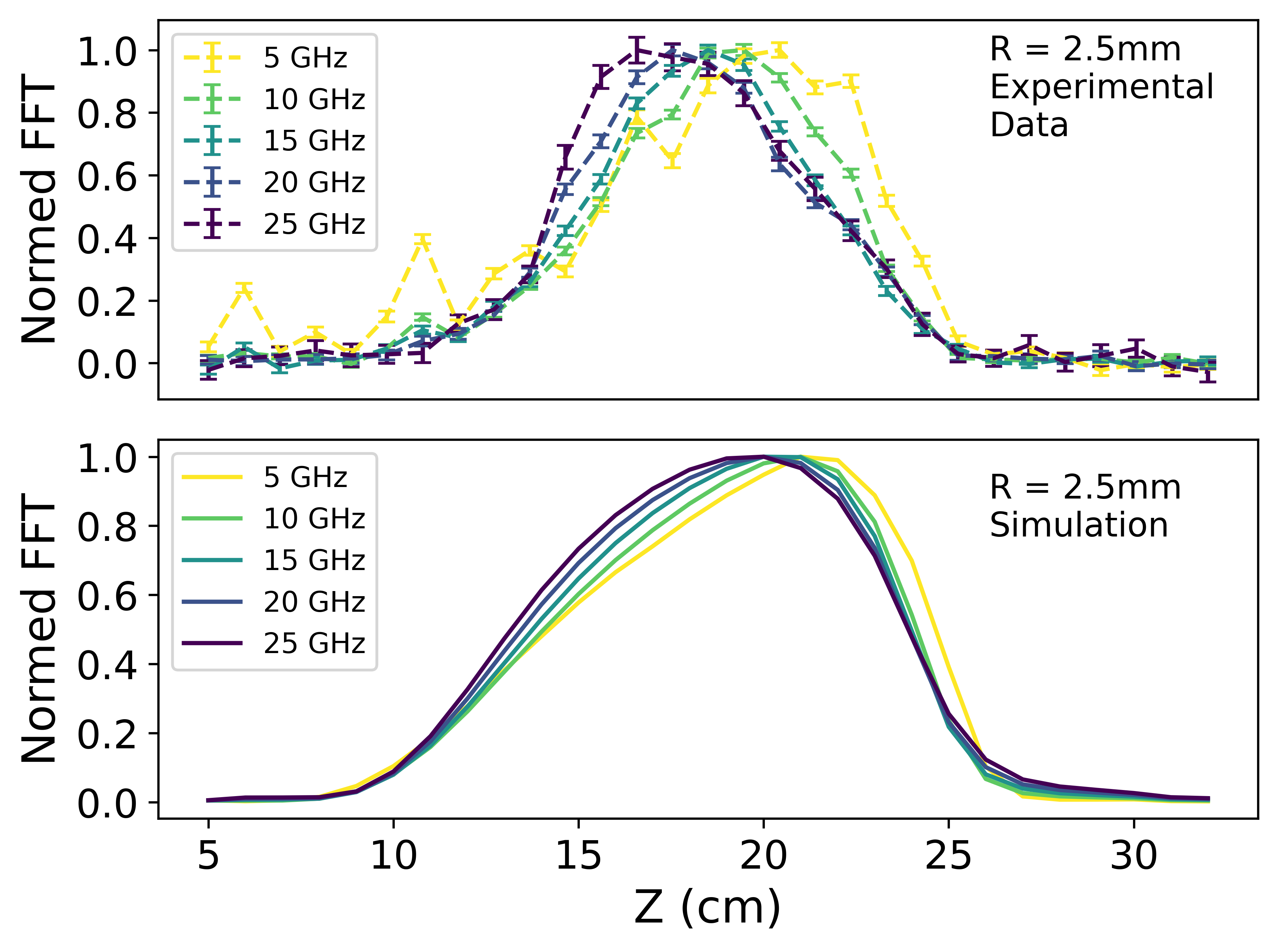}
  \caption{Normalized traces of measured electric field strength in 
  5 different frequency bins as a function of longitudinal 
  position $z$, for both the experimental data and the 
  corresponding simulation shown in Fig. \ref{sim_pics}.
  In both cases the higher frequency components of the wave 
  reach their maximums at earlier $\hat{z}$ positions, 
  and the lowest frequencies reach their maximums last.
  The experimental data also becomes successively more 
  noisy at lower frequencies, which we suspect is due to 
  a combination of phase velocity drop off at low frequencies, 
  and focusing and defocusing effects modulating the plasma 
  column density.
  }
  \label{drift_exp_sim_combo}
\end{figure}

Fig. \ref{fft_plots} shows the FFTs taken at 4 radial separations ranging 
from $r=0.5$ cm to $r=1.5$ cm, taken near the peak at $z = 18$ cm, 
as well as the noise signal recorded when 
the laser pulse is blocked at the focusing lens and no plasma forms
(note that each curve is the average of 50 individual shots). 
Although the data is uncalibrated the ultra-broadband nature of the 
near field waves is clear, (as can also be seen in the original 
time series data), and the overall profile resembles the far field RF 
recorded in \cite{englesbe2018gas}.

We next average the radial FFT data over large frequency bins so 
that we can compare with theory. In general the radial fall off 
is very well fit by the expected Hankel function dependance in (\ref{eqn_Sommerfeld_Goubau}), 
and is much better than naive polynomial fits.  
Fig \ref{hankel_fits} shows the single best fit that occurs at 14 GHz 
(which is also about the frequency of largest signal - see Fig. \ref{fft_plots}), 
with an $R^2$ value of 0.9993.
All of the frequency bins are best fit by the Hankel function dependence, 
with typical $R^2$ values of about 0.99.
The radial dependence of simulated SWs (see Fig. \ref{sim_pics}) is also very well fit 
by a Hankel function, with $R^2$ values better than 0.9999.

We note that the Hankel function fit underpredicts the measured 
data starting at about 2.5 cm in Fig. \ref{hankel_fits}, 
and this trend is seen in the other frequency bins. 
In simulation some free radiation is also excited by 
the radial $J_r$ currents, and the amplitude of this 
radiation will fall off more slowly than the 
Hankel function at large radius ($r > r_{\nouter}$).
We thus suspect that we are also detecting this free radiation, 
although the overall amplitude of the measured signal 
is small at this larger radius.  

Given the Hankel function fits we can also extract the 
fit values for the outer length scale $r_{\nouter}$ and 
compare with theory. We find that the experimental 
$r_{\nouter}$ values match well with an electron density 
of $n_e = 3\times10^{22}$ m${}^{-3}$ (see Fig. \ref{theory_pics}), 
with an experimental fit $r_{\nouter} = 0.75$ cm 
at 15 GHz being close to the theory value given by 
(\ref{iterative_solve}).
However the slope of the experimental values as a function of 
frequency is lower than the theoretical curve, as can be 
seen in Fig. \ref{hankel_fits}.  

We next consider the longitudinal variation in the measured D-dot signal. 
Fig. \ref{drift_exp_sim_combo} shows the amplitude as 
a function of $z$ for 5 frequency bins ranging from 5 GHz to 25 GHz, 
as measured at $r = 2.5$mm in both the experimental 
and Drude simulation data (see Fig. \ref{sim_pics}). 
As a D-dot calibration is not available the different frequencies have all been 
normalized to 1 for comparison purposes.
We note that there is a consistent drift of the peaks of the 
different frequencies to larger $z$ as the frequency decreases 
in both the experimental data and in the simulations. 

This longitudinal drift towards lower frequencies 
is consistent with the theoretical expectations shown in 
Fig. \ref{theory_pics}: from higher frequencies down to about 
5 GHz the phase velocity of the SPPs is approaching $c$, 
and thus the lower frequency surface waves have better phase 
coherence with the $J_r$ current source. 
The lower frequency waves also decay more 
slowly, helping them to build to a later peak.
At even lower frequencies below 5 GHz the phase velocity rapidly drops 
(depending sensitively on the electron density), 
and the waves no longer co-propagate efficiently with 
$J_r$, and the amplitude of the SPPs falls off quickly.
We note that the experimental curves become much noiser
below 5 GHz, which we suspect is due to density 
fluctuations along the plasma column due to 
focusing and de-focusing effects, and which the 
lower frequencies are more sensitive to.
At the larger $z$ values towards the end of the plasma column
the electron density $n_e$ drops off, which causes the waves to both decay 
more quickly and detach from the plasma column, leading to a collapse 
of the signal.

Inspired by exploratory simulations we argued in \cite{garrett2021generation} that 
SPPs are both generated during femtosecond filamentation, and grow in strength due to coherent 
copropagation near the speed of light with the plasma wake currents that excite them 
along the leading 
edge of the plasma column.
In this paper we present new near field data 
that is in excellent agreement with this theory.
We find that the radial profile of the D-dot measurements is 
in very good agreement with the expected Hankel function 
dependence from the Sommerfeld-Goubau solution, 
and the characteristic outer length scale is well 
matched by a plasma with density $n_e = 3\times10^{22}$ m${}^{-3}$. 
In turn the longitudinal frequency drift data seen in the data is in 
good agreement with both theoretical expectations for the phase 
velocity and decay length (\ref{iterative_solve}), and with 
new simulations (Fig. \ref{sim_pics}) 
that include variations in plasma density \cite{janicek2023measuring}.
The near field measurement of these surface waves thus provides a 
powerful new scientific tool for chirped pulse filamentation research.

\begin{acknowledgments}

\textit{Acknowledgments}. The authors thank the Air Force Office of Scientific Research (AFOSR) for support
via Laboratory Task No. FA9550-24RDCOR002.
This work was supported in part by high-performance computer time
and resources from the DoD High Performance Computing Modernization Program.
Approved for public release; distribution is unlimited. Public Affairs
release approval No. AFRL-2024-6758.

\end{acknowledgments}

\bibliography{../bib_rf,../surface_waves}

\end{document}